%% file: main.tex
\documentclass{aa}  

\usepackage[OT1]{fontenc} %%%%% for \AE compatibility
\usepackage{graphicx}
\usepackage{microtype} %%%%% typographical refinements
\usepackage{booktabs} %%%%% quality tables
\usepackage[pdfauthor={Nikolas Frediani, Daniel Gruen, Luca Tortorelli, Jamie McCullough},
            pdftitle={Data-driven Galaxy Population Prior for Photometric Redshifts},]{hyperref}
\graphicspath{ {figures/} } %%%%% figurepath
\usepackage[group-separator={,}, group-digits=integer]{siunitx} %%%%% SI unit and number formatting
\DeclareSIUnit\angstrom{\text{\AA}}
 
\DeclareMathOperator*{\argmin}{argmin} 
\usepackage[capitalise]{cleveref} %%%%% clever references
\usepackage{txfonts}
\usepackage{lipsum}
\usepackage{subcaption}         % necessary for continued figures, example in section 3
\usepackage{lscape}             % to rotate a single page table, example in appendix.
\usepackage{placeins}           % useful with \FloatBarrier, to keep 
\usepackage[dvipsnames]{xcolor}
\usepackage[normalem]{ulem}

\begin{document}

%%%%%%%%%%%%%%%%%%%%%%%%%%%%%%%%%%%%%%%%
% if you use custom commands in your title,
% ensure to check your title when submitting!
%%%%%%%%%%%%%%%%%%%%%%%%%%%%%%%%%%%%%%%%
   \title{Data-driven Galaxy Population Prior for Photometric Redshifts}

   \subtitle{}

%%%%%%%%%%%%%%%%%%%%%%%%%%%%%%%%%%%%%%%%
% Please separate each author with the \and command
%
% Please do not include ORCIDs next to author names.
% Only ORCIDs authenticated by individual authors in EDPS
% editorial system will be taken into account.
% ORCIDs included here will be removed.
%%%%%%%%%%%%%%%%%%%%%%%%%%%%%%%%%%%%%%%%

   \author{N. Frediani\inst{1}\fnmsep\thanks{Corresponding author: Nikolas.Frediani@lmu.de}
        \and D. Gruen\inst{1}\fnmsep\inst{2}
        \and L. Tortorelli\inst{1}
        \and J. McCullough\inst{3}
        }

   \institute{Universitäts-Sternwarte, Fakultät für Physik, Ludwig-Maximilians Universität München, Scheinerstr. 1, 81679 München, Germany
   \and Excellence Cluster ORIGINS, Boltzmannstr. 2, 85748 Garching, Germany
   \and Department of Astrophysical Sciences, Princeton University, Princeton
NJ 08544, USA}

   \date{Received Month 0X, 20XX}

% \abstract{}{}{}{}{}
% 5 {} token are mandatory
 
  \abstract
   {Stage-IV cosmological surveys rely on well-characterised photometric redshift distributions for constraining cosmological models, yet the projected weak lensing requirements are about an order of magnitude more accurate than current state-of-the-art methods. Any photometric redshift inference depends - explicitly or implicitly - on a prior over galaxy SEDs, luminosities, and redshifts.
   }
   {In this paper, we develop a purely data-driven approach for modelling the galaxy population prior based on learning the distribution of galaxy SEDs and evaluate its induced systematic uncertainties for photometric redshift calibration.
   Under controlled conditions, we build a template-free model for galaxy SEDs with minimal physical assumptions and assess how well it can reproduce the colour-redshift relation.
   } 
   {We train a generative model on a realistic population of noisy mock galaxy spectra, simulated using the \textsc{GalSBI-SPS} galaxy population model, to learn the joint distribution of intrinsic spectra, luminosities, and redshifts. 
   We employ a probabilistic autoencoder that compresses spectra into a low-dimensional latent space and performs neural density estimation with a normalising flow. 
   }
   {The autoencoder can reconstruct the shape of galaxy spectra with Gaussian noise of standard deviation $\sigma$ to $\sim0.1\,\sigma$ of the ground truth, thereby building a model for noiseless SEDs from noisy spectra only. 
   We construct colour-selected tomographic bins with a self-organising map and compare the predicted mean redshift.
   We find that deviations in each bin are smaller than the per-mille Stage-IV requirements, with $| \Delta \langle z \rangle |\lesssim 0.0007 (1+z)$.
   }
   {This work serves as a proof of concept that generative models can learn the prior of galaxy observations accurately enough for upcoming surveys under controlled conditions.
   }

    % galaxy population -- spectroscopic surveys -- photometric redshifts -- generative model

   \maketitle
\nolinenumbers %%%%% remove line numbers

%\tocite
%\nf{comments}\dgr{edits}{edited}

%fontsize: 
%\makeatletter
%\f@size pt
%\makeatother\\
%textwidth: \the\textwidth\\
%linewidth: \the\linewidth\\
%hsize: \the\hsize\\

%%%%%%%%%%%%%%%%%%%%%%%%%%%%%%%%%%%%%%%%%%%%%%%%%%%%%%%%%%%%%%

\section{Introduction}
\label{ch:intro}
\input{sections/1Introduction}

\section{Methods}
\label{ch:methods}
\input{sections/2Methods}

\section{Results}
\label{ch:results}
\input{sections/3Results}

\section{Discussion}
\label{ch:discussion}
\input{sections/4Discussion}

\section{Conclusion}
\label{ch:conclusion}
\input{sections/5Conclusion}

%%%%%%%%%%%%%%%%%%%%%%%%%%%%%%%%%%%%%%%%%%%%%%%%%%%%%%%%%%%%%%
\begin{acknowledgements}
NF acknowledge funding via the SciFM consortium (05D25WM2) funded by the German Federal Ministry of Research, Technology, and Space (BMFTR) in the ErUM-Data action plan. Funded by the Deutsche Forschungsgemeinschaft (DFG, German Research Foundation) under Germany's Excellence Strategy – EXC-2094/2 – 390783311.
\end{acknowledgements}

% Please note that we have included the references below in
% order to compile the document, but we ask you to:

\bibliographystyle{bibtex/aa.bst} % style aa.bst
\bibliography{bib.bib} % your references Yourfile.bib

%%%%%%%%%%%%%%%%%%%%%%%%%%%%%%%%%%%%%%%%%%%%%%%%%%%%%%%%%%%%%%%
% Appendices must be placed after   \end{thebibliography}
% They will be placed automatically on a new page.
%%%%%%%%%%%%%%%%%%%%%%%%%%%%%%%%%%%%%%%%%%%%%%%%%%%%%%%%%%%%%%%
\begin{appendix}
\nolinenumbers %%%%% remove line numbers
%%%%%%%%%%%%%%%%%%%%%%%%%%%%%%%%%%%%%%%%%%%%%%%%%%%%%%%%%%%%%%%
% In the PDF output, floats should be placed
% under their own appendix, not before the title, nor after the
% title of the next appendix.

% In short appendices, onecolumn floats (\figure*
% or \table*) will generate a blank page.
% To prevent this behaviour, a few examples are provided here. 

% In case you have a lot of floating objects for little text and the 
% LaTeX engine moves the floats away from their context, the command
% \FloatBarrier of the “placeins” package will empty the
% float buffer and place all stored floats in the continuity.

% If you still encounter problems with wide floats placement,
% just use the onecolumn environment throughout the appendices.
%%%%%%%%%%%%%%%%%%%%%%%%%%%%%%%%%%%%%%%%%%%%%%%%%%%%%%%%%%%%%%%

% \crefalias{section}{appendix}
\input{sections/Appendix}

\end{appendix}
\end{document}

%% file: sections/1Introduction.tex
% gen 4 galaxy surveys rely on photo-z characterisation
The upcoming Stage-IV of galaxy surveys (e.g. Euclid \citep{euclidcollaborationEuclidOverviewEuclid2025}, LSST \citep{ivezicLSSTScienceDrivers2019}, and Roman \citep{akesonWideFieldInfrared2019}) will provide data of unprecedented quality and scale, opening the door for precision measurements of cosmological parameters, but at the same time demanding increased accuracy in the control of systematics, which in turn requires increased accuracy of the physical models and techniques used to analyse the data.
In particular, the challenge of calibrating photometric redshifts (photo-$z$s) has emerged as a major frontier that will need an order of magnitude improvement in the coming years. 
It is likely to be the limiting systematic for Stage-IV cosmology \citep{dalalHyperSuprimeCamYear2023,liHyperSuprimeCamYear2023}.
\citet{newmanPhotometricRedshiftsNextGeneration2022} distinguish between two goals: 'precision', i.e. estimating photo-$z$ point estimates or posteriors for individual galaxies, and 'characterisation', i.e. obtaining the redshift distribution $n(z)$ of populations of galaxies. Because the weak lensing (WL) signal is mostly sensitive to the mean redshift $\langle z \rangle$ of source galaxies in broad tomographic bins, the latter is of special interest for cosmology \citep{amonDarkEnergySurvey2022,corderoDarkEnergySurvey2022}.

% LSST and Euclid requirements, Stage III SOTA
The Legacy Survey of Space and Time (LSST) \citep{ivezicLSSTScienceDrivers2019} Dark Energy Science Collaboration demands that the mean redshift of a tomographic bin be known better than $| \Delta \langle z \rangle | \leq 0.002(1+z)$ for Y1 weak lensing analysis, which tightens further to $| \Delta \langle z \rangle | \leq 0.001(1+z)$ for Y10 \citep{collaborationLSSTDarkEnergy2021}. 
Similarly, Euclid \citep{euclidcollaborationEuclidOverviewEuclid2025} quotes an upper bound of $| \Delta \langle z \rangle | \leq 0.002(1+z)$ of the systematic uncertainty on mean redshift \citep{laureijsEuclidDefinitionStudy2011}.
However, this is about an order of magnitude below current state-of-the-art photo-$z$ methods based on redshift calibration, with Stage-III redshift distributions being known down to $| \Delta \langle z \rangle | \gtrsim 0.01 (1+z)$ \citep{yinDarkEnergySurvey2025,wrightKiDSLegacyRedshiftDistributions2025}. 

% photo-z calibration
Most methods for assigning redshifts with photometry rely on a set of spectroscopic observations for calibrating the colour-redshift-relation \citep{mastersMAPPINGGALAXYCOLOR2015}. This depends on obtaining a spectroscopic calibration sample that is representative of the photometric sample in colour-magnitude space. While this is limited due to selection effects and the cost of deep spectroscopy, the Complete-Calibration-of-the-Colour-Redshift-Relation (C3R2) survey \citep{mastersCompleteCalibrationColor2017, mastersCompleteCalibrationColor2019, stanfordEuclidPreparationXIV2021} aims to systematically fill the galaxy colour space with spectroscopic observations. 
Recently, advances are being made with the Dark Energy Spectroscopic instrument (DESI) \citep{j.mcculloughDESICompleteCalibration2024, deyDeepSpectroscopyDESI2026, blancoPowerDESIPhotometric2025} and 4-metre Multi-Object Spectroscopic Telescope (4MOST) \citep{gruen4MOSTCompleteCalibration2023}.

% photo-z methods
Photo-$z$ estimation can be understood in a Bayesian sense, where the redshift posterior is obtained by integrating the likelihood and a prior over galaxy types \citep{newmanPhotometricRedshiftsNextGeneration2022}:

\begin{equation}
    p(z|f_i) \propto \int p(f_i|\mathcal{S}, z) p(\mathcal{S}, z) d\mathcal{S}
\end{equation}

with photometric fluxes $f_i$ and a - continuous or discrete - set of spectral types denoted by $\mathcal{S}$.
The right-hand side could potentially also depend on additional variables, such as observational properties of galaxies and cosmology, which we neglect here for simplicity.

In this framework, the challenge comes down to how to model the prior of galaxy observations $p(\mathcal{S},z)$.
Colour-based photo-$z$ methods can broadly be categorised in three approaches that differ in how they model this prior: template-based methods, empirical methods, and hybrid/forward-modelling approaches. 
The former relies on a set of pre-defined spectral templates, while data-driven methods inherit the prior from a training sample. Finally, hybrid and generative approaches have different ways of incorporating their priors, but typically also fall back to either of the previous two assumptions \citep[e.g.][]{salvatoManyFlavoursPhotometric2018, newmanPhotometricRedshiftsNextGeneration2022}.

% template methods
Template-based methods include \textsc{BPZ} \citep{benitezBayesianPhotometricRedshift2000}, \textsc{LePhare} \citep{arnoutsMeasuringModellingRedshift1999,ilbertAccuratePhotometricRedshifts2006}, and \textsc{EAZY} \citep{brammerEAZYFastPublic2008}. They differ mainly in what template set is used and their treatment of the prior on the templates, as well as whether or not they report a full posterior or just point estimates.
While these codes are still prominently used as baselines in the literature, they are fundamentally limited by our ability to model the galaxy population precisely. Their accuracy suffers if the set of templates is incomplete or if the prior is not calibrated well enough.

% empirical methods
Empirical methods learn the colour-redshift-relation with machine learning algorithms.
The approaches are typically either based on regression (using e.g., Neural Networks \citep{collisterANNzEstimatingPhotometric2004, sadehANNz2PhotometricRedshift2016, jonesImprovingPhotometricRedshift2024}, random forests \citep{carrascokindTPZPhotometricRedshift2013}, or Gaussian processes \citep{almosallamGPzNonstationarySparse2016, gomesImprovingPhotometricRedshift2018}) 
or similarity (using e.g. k-nearest-neighbours \citet{devicenteDNFGalaxyPhotometric2016, grahamPhotometricRedshiftsLSST2017}, self-organising maps \citet{mastersMAPPINGGALAXYCOLOR2015}, or UMAP \citet{ashmeadOptimizingPhotometricRedshift2025}). Some works expand the informativeness of the data by considering not only fluxes, but images of galaxies \citep{schuldtPhotometricRedshiftEstimation2021, deyPhotometricRedshiftsSDSS2022, fathkouhiAstroMAERedshiftPrediction2024, merzDeepDISCphotozDeepLearningBased2025, luoColorsProbingRedshifts2026}.
See \citet{tortorelliMachineLearningTechniques2026} and references therein for a review of the landscape of AI approaches for photometric redshifts.

% SOM
Of special interest for this work are self-organising map (SOM)-based approaches \citep{kohonenSelfOrganizingMaps2001, mastersMAPPINGGALAXYCOLOR2015}, that bin galaxies by colour similarity. These give asymptotically correct characterisation in the limit of a representative and abundant calibration sample. For this reason, they were used by both the DES and KiDS WL analyses (e.g., \citealt{yinDarkEnergySurvey2025,wrightKiDSLegacyRedshiftDistributions2025}). The SOM algorithm is further described in \cref{ch:photo-z}.

% (dis)advantages of empirical methods
These empirical methods can circumvent the limitations of templates and have no \textit{a priori} theoretical limit on their performance, scaling with more training data.
However, they rely heavily on deep and abundant training samples that are representative of the variety of colours of the weak lensing sample.
Moreover, the aforementioned approaches generally treat galaxy colours as independent: they learn the mapping from colours to redshift without taking correlations between galaxies of different colours into account. However, galaxies with intrinsically similar SEDs can appear at different redshifts, where they will be observed with different colours, while they are physically correlated through the intrinsic spectrum. If this information is not exploited, that will contribute to the limitations of many empirical methods.

% forward modelling populations
Another approach of obtaining the distribution of galaxies is in full forward-modelling of the population. A wide variety of models have been developed based on different prescriptions, from semi-analytic to parametric and data-driven models. % Synthesizer, GalSyn, Diffsky, SHARK
When constrained to data, these models have a wide range of applications, from forward modelling survey observations to informing the physical properties of the galaxy population itself. Most importantly for cosmology, they allow to constrain redshift distributions which can then be used either as the prior for photo-$z$ inference of individual objects or directly in a cosmological analysis.

% GalSBI
An example of this approach is the phenomenological population model \textsc{GalSBI} \citep{fischbacherGalSBIPhenomenologicalGalaxy2025, fischbacherGalSBIForwardModelling2026} and its stellar population synthesis (SPS)-based counterpart \textsc{GalSBI-SPS} \citep{tortorelliGalSBISPSStellarPopulation2025}, based on parametric descriptions of population properties and their relations. Galaxy observations are sampled conditionally from the underlying properties, allowing for fast and realistic forward-modelling of populations with physically motivated functional forms of galaxy abundances and evolution. They can be constrained to match observational data, yielding posteriors of model parameters for which the resulting population matches observation.

% pop-cosmos
Alternative population models targeting redshift distributions include the more data-driven \textsc{pop-cosmos} \citep{alsingForwardModelingGalaxy2023, thorpPopcosmosScaleableInference2024, leistedtPopcosmosForwardModeling2026, halderPopcosmosRedshiftsPhysical2026}, which samples directly from the joint distribution of galaxy properties and forward-models broad-band fluxes.
The prior is constrained by minimizing the distance of a generated catalogue to COSMOS data and can be used to inform photo-$z$ inference.
A similar approaches include e.g. \textsc{PopSED} \citep{liLiPopSEDPopulationlevelInference2024}, a flow model for inferring population-level properties from photometry based on SPS, and \textsc{PZFlow} \citep{crenshawProbabilisticForwardModeling2024}, a density estimation on the catalogue level.

% problems of forward models
However, when considering forward-modelling priors solely for the purpose of redshift calibration, population models suffer ultimately suffer from similar weaknesses as other physical photo-$z$ approaches, such as assumptions and choices in the forward-model (e.g. in the SPS-model), or on the functional form of property relations. Thus, it is not evident if one of these models can be tuned finely enough for Stage-IV. Moreover, because of their complexity and degeneracies and without access to the true underlying population it is unclear how to validate and optimise any given model. For example, \citet{tortorelliImpactStellarPopulation2024} have shown that choices in the SPS prescription can significantly bias the resulting mean redshift of a sample, which is therefore a concern for all models relying on SPS codes.

% recent related works
Finally, several hybrid approaches have been proposed for photo-$z$ inference that combine several aspects of the aforementioned categories. 
E.g. \citet{federProbabilisticAutoencoderGalaxy2026} use a probabilistic autoencoder to model the galaxy prior, similar to this work, and rely on spectral simulations and an explicit, externally provided prior.
\citet{guoJointProbabilisticInference2026} built a model for inferring redshifts that is similarly based on latent density estimation with a diffusion model on \textsc{SPENDER} \citep{melchiorAutoencodingGalaxySpectra2023} redshift-invariant representations of DESI spectra.

% not good enough!
However, despite many advances in algorithms and a broad spectrum of approaches, no unique, clear path forward has yet emerged. No method has quantitatively proven that it can reach the stringent Stage-IV requirements necessary for next-generation WL. 
A benchmark by \citet{schmidtEvaluationProbabilisticPhotometric2020} tested many codes available at the time for LSST photometric redshifts and showed that even under ideal conditions, that is with a representative spectroscopic sample and appropriate templates, current methods fail to achieve the required characterisation accuracy. 
Newer approaches struggle to make significant leaps in performance. Additionally, many papers focus on evaluating photo-$z$ precision, that is on comparing point-estimates or posteriors on individual galaxies, and omit evaluation on the population level characterisation.

% 3 requirements on prior models that we need to fulfil
We can break down the ways in which models fail to inform the necessary qualities that future approaches will have to fulfil: 
1. Templates -- and any assumption on the population -- are not going to be good enough approximations because that would require obtaining representative galaxy SEDs down to the photometric depth of Stage IV surveys. Hence, a good prior model should be template-free with no or minimal assumptions. 
2. Existing data-driven models appear to not exploit the available data to their full potential. A good prior model should incorporate the concept of intrinsic rest frame spectra, thereby explicitly using the fact that galaxies of similar spectral type can appear at different redshifts, instead of treating colours as independent.
3. Finally, we need to define an appropriate metric of success, as the usual per-galaxy metrics of scatter, outlier-fraction, posterior coverage, etc. are not appropriate for the cosmology goal. That means that the prior needs to match the Stage-IV per-mille requirements on the mean redshift of colour-selected tomographic bins.  

% our solution: generative model of restfr
To address these issues, we propose a data-driven approach for modelling the galaxy population prior. We use a generative model for learning the distribution of normalised rest frame spectra, redshifts and luminosities: $p(\mathcal{S},z,L)$. 
Advances in machine learning and generative modelling have supplied us with powerful techniques that allow us to realise a generative model as an expressive and flexible instance of the galaxy prior.

% why our approach is good
The model is agnostic to the physics of the galaxy population. As outlined before, we explicitly want the model to be informed only by data with no outside information. 
For this reason, we only make two elemental assumptions, which are (1) how intrinsic SEDs evolve under redshift (by stretching the observed wavelength) and (2) that the same SED can be realized by galaxies of a range of luminosities. 
By operating in the rest frame, we build a model for intrinsic galaxy types that explicitly translates across redshifts.

% scope and goals of this paper
In this paper we explore whether a spectral generative model can learn the galaxy prior to the desired precision. As a first step and proof of concept, we focus on evaluating the learned colour-redshift-relation. To this end, we assume a representative spectroscopic sample and investigate how accurate the model can reproduce the mean redshift of colour-selected samples.

% validate on ground truth
Because the ground truth of intrinsic spectra is not accessible in observations, we first validate the generative approach on simulations. By training on a realistic but known synthetic population of galaxies, we are able to quantify exactly how reliable we recover the intrinsic population. Since the machine learning architecture is agnostic to the physics and details of the population, we can expect the results to generalise to data, if the simulations are realistic enough. Note that we do not require the population itself to match data at this point, this work only serves as a proof-of-concept and shows whether generative models can achieve the desired accuracy under simplifying assumptions.

% we evaluate on colour accuracy and photo-zs
We evaluate performance of the generative model not only by comparing the learned distributions, but also investigate the end-to-end impact of the trained model for a realistic WL photo-$z$ calibration. We split the sample into SOM-based tomographic bins, and compare the redshift distributions of the simulated and a generated sample. 

% structure
This paper is structured as follows:
In \cref{ch:methods} we introduce the simulations, machine learning framework, and photo-$z$ calibration pipeline we use.
\Cref{ch:results} shows the performance of the trained networks individually, as well as the overall impact on redshift estimation.
Finally, we discuss the impact and limitations of this work in \cref{ch:discussion} and conclude in \cref{ch:conclusion}.

%% file: sections/2Methods.tex
% structure
Here we describe the methods we use to build and test the generative model. 
\Cref{ch:simulations} details the simulations of our training data set. 
The machine learning architecture for the generative model itself is explained in \cref{ch:generative_model}.
We describe our synthetic photometry in \cref{ch:photometry} and photo-$z$ calibration prescription in \cref{ch:photo-z}.

\subsection{Simulations}
\label{ch:simulations}

% we use simulations to validate the principle, but are ultimately agnostic
We simulate a dataset of galaxy properties and spectra, which represents the underlying true population in the context of this work.
We emphasise that the specific model choice is not fundamental to the method as long as the resulting spectra show sufficient diversity. The machine learning algorithm presented is agnostic to the details of the simulations, such as spectral templates used, population distributions, and survey- or instrument-specific effects. Ultimately, we aim for a model that is trained on observational data directly. Here we use the simulations as a means to validate the generative model and assess it's ability to recover the true distribution from noisy observations.

% GalSBI-SPS to generate 1M dataset
We use the \textsc{GalSBI-SPS} \citep{tortorelliGalSBISPSStellarPopulation2025} galaxy population model to generate the training dataset. \textsc{GalSBI-SPS} is a SPS-based model that samples galaxy properties from parametric functions conditioned on stellar mass and redshift, and uses the SPS code \textsc{Prospect} \citep{robothamProSpectGeneratingSpectral2020} to simulate spectra.
We generate \num{2305000} galaxies with a magnitude cut of $m_z \leq 22$ and a redshift cut of $z_\mathrm{max} = 1.55$. 
These cuts were chosen to approximate a 4MOST-like sample, which can be used to calibrate photo-$z$s for surveys such as Euclid.
The magnitude cut matches the upcoming 4C3R2 (deep) survey depth \citep{gruen4MOSTCompleteCalibration2023}, while $z_\mathrm{max}$ is the redshift at which the [OII] emission-line doublet, a critical feature for determining spectroscopic redshifts, leaves the observed optical range of the 4MOST spectrograph \citep{dejong4MOST4metreMultiObject2014}. 
While this does not cover the full depth of a Stage-IV survey, the effective galaxy samples used in weak-lensing and related large-scale-structure analyses can be considerably shallower than the imaging depth, making the adopted selection relevant to the regime considered here \citep{porredonDarkEnergySurvey2022,deroseSteelingWeakLensing2026}.
We discuss the assumption of the representativeness of the spectral data in \cref{ch:discussion}.

% train-test-split
From the full dataset, we randomly select \num{1000000} spectra for training, with \num{1000000} and \num{305000} spectra reserved for testing and validation, respectively. 
The $10^6$ objects in the training set motivated by the order of magnitude of representative spectra that can be obtained with current and upcoming surveys. The test set size is deliberately chosen to be of the same order as the training set to allow for high-statistics evaluation of the generative model, as required by the precision nature of the application.

% wavelength coverage
For the photo-$z$ calibration we apply in \cref{ch:photo-z_results}, we need to cover the wavelength range from observed \textit{u}- to \textit{$K_s$}-band, which corresponds to a total wavelength coverage of $\qty{1000}{\angstrom} \leq \lambda \leq \qty{25000}{\angstrom}$ in rest frame for galaxies in $0<z<1.55$.
While this coverage is idealised in that currently no multiplexed instrument can simultaneously obtain well measured spectra over such a wide wavelength range, we make this simplifying assumption as a proof of concept (see \cref{ch:discussion}).
We re-bin each rest frame spectrum to a fixed wavelength grid of 2000 logarithmic bins, which corresponds to a resolution of $R \sim 620$.
While some physical information is lost by re-binning to this lower resolution, mostly in the shape of line-features, we do the re-binning in a flux-conserving way. Thus colours are not affected, which is the primary interest of this study.
Note also that this binning is arbitrary and was chosen mostly for computational efficiency. We observed the autoencoder to scale well with the spectral input dimension, so this architecture can easily be adapted to other wavelength grids, such as higher resolution spectra or different wavelength ranges.

% preprocessing & normalisation
Before the spectra are fed to the network, they are preprocessed for numerical stability. 
We define the spectrum normalisation $L^*$ as the median flux density:
\begin{equation}
    L^* = \mathrm{median}(F_\lambda) .
\end{equation}

Note that this depends on the wavelength grid we chose, and thus does not directly correspond median physical flux density due to the logarithmic binning, but is an equivalent encoding of the effective luminosity or amplitude of a spectrum.
We normalise each SED by dividing by it's median flux density:

\begin{equation}
    f_\lambda = \frac{F_\lambda}{L^*}
\end{equation}

% noise
Finally, we add artificial noise to the intrinsic spectra to simulate more realistic observational conditions and investigate the de-noising capabilities of our approach. For simplicity we apply uncorrelated noise with a constant amplitude proportional to the flux value in each wavelength bin. 

\begin{equation}
    \tilde{f_\lambda} = f_\lambda \cdot (1 + \eta); \quad\quad \eta \sim \mathcal{G}(0,\Sigma)
\end{equation}

In \cref{ch:appendix_scaling} we investigate how our model scales when we increase the noise level $\Sigma$. Throughout the remainder of this paper, we adopt a noise level of $10\%$ as our fiducial model. 
We note that noise in real spectra is usually non-gaussian, can vary as a function of wavelength and observing conditions, and heterogeneous across the dataset (see \cref{ch:discussion} for discussion).

\subsection{Generative Model}
\label{ch:generative_model}

% PAE schematic
\begin{figure}
    \centering
    \includegraphics[width=\linewidth]{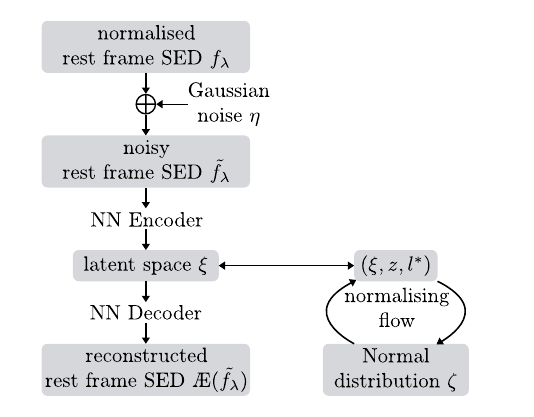}
    \caption{A schematic representation of the probabilistic autoencoder (PAE). Inputs are compressed with a NN, and reconstructed with a mirrored architecture. A normalising flow transforms between the latent space distribution and a simple Gaussian.}
    \label{fig:PAEschematic}
\end{figure}

% PAE
In this paper, we train a probabilistic autoencoder (PAE) \citep{bohmProbabilisticAutoencoder2022} as a generative model for galaxy SEDs.
The PAE is a combination of two neural networks (NNs): a classic autoencoder (AE) that learns a low-dimensional representation of the data which can be decoded back to full spectra in physical space. 
%compresses high-dimensional spectra into a low-dimensional latent space and then reconstructs the original inputs. 
This is followed up by a density estimation of the latent space with a normalising flow, which enables sampling from the latent space.
\Cref{fig:PAEschematic} shows a schematic representation of the network architecture.

% PAE advantages
The PAE architecture disentangles different aspects of the data processing: the AE finds compressed representations of high-dimensional SEDs and removes noise, while the flow learns the distribution of the more manageable latents. 
Density estimation is inherently plagued by the curse of dimensionality, i.e. a high number density is required to model the PDF accurately. Therefore te AE is crucial for our goal of obtaining a precisely calibrated prior.
Moreover, the PAE is a conceptually very straight-forward and modular framework: both networks are trained separately and can be optimized, adapted, or upgraded individually to improve the overall accuracy or address specific aspects of the data. 
We intentionally choose simple NNs at this point to demonstrate the feasibility of our data-driven generative approach to photo-$z$s and leave implementing more elaborate architectures to future work.

\subsubsection{Autoencoder}
\label{ch:ae}

% AE introduction
AEs \citep[e.g.][]{hintonReducingDimensionalityData2006, liComprehensiveSurveyDesign2023, Bank2023} are neural networks that compress inputs to a low-dimensional latent space and try to reconstruct the original inputs. These latents will be an effective description of the full spectra and encode the relevant physical information contained in the data.
The AE directly addresses one observational effect: the latent space acts as an information bottleneck and will thus converge to only encode the information that is common across the dataset, while ignoring uninformative features such as uncorrelated noise. Thus, the AE will implicitly denoise the data, which allows us to reconstruct approximations of noiseless spectra when only noisy spectra are available for training.
AEs have found widespread application in processing astrophysical galaxy spectra for compression/representation learning \citep[e.g.][]{portilloDimensionalityReductionSDSS2020, melchiorAutoencodingGalaxySpectra2023, guoJointProbabilisticInference2026, federProbabilisticAutoencoderGalaxy2026}, denoising \citep{scourfieldDenoisingGalaxyOptical2023} and anomaly detection \citep{bohmFastEfficientIdentification2023, liangAutoencodingGalaxySpectra2023, liangOutlierDetectionDESI2023, nicolaouIdentifyingAnomalousDESI2026}.

% latent space size
Galaxy spectra in data space are very high-dimensional objects with thousands of entries, depending on the wavelength grid. However, their inherent dimensionality is much lower.
Previous empirical studies have found that $\sim$\,\numrange{4}{10} numbers are enough to describe the diversity of spectra and achieve good reconstructions \citep{portilloDimensionalityReductionSDSS2020, melchiorAutoencodingGalaxySpectra2023, scourfieldDenoisingGalaxyOptical2023, federProbabilisticAutoencoderGalaxy2026}. After that, the reconstruction accuracy plateaus, and adding more latent dimensions gives diminishing returns on the informativeness of the representation.
Based on our own scaling tests we adopt a latent dimension of 6 (see \cref{ch:appendix_scaling}).

% MLP NN
We adopt a simple deep NN (MLP) architecture, trained by optimizing the reconstruction loss given by 

\begin{equation}
    \mathcal{L}_\mathrm{reco}= \frac{1}{N} \sum_{i=1}^{N} \Biggl\langle \frac{( \AE (f_{\lambda,i})-f_{\lambda,i})^2}{\sigma_{\lambda,i}^2} \Biggr\rangle_\lambda , 
\end{equation}

with N spectra $f_{\lambda,i}$ with uncertainties $\sigma_{\lambda,i}$, and denoting the function given by the NN as $\AE(f_\lambda)$. The re-weighting of the MSE loss by the inverse variance naturally incorporates varying SNR and accounts for heterogeneities in the data.
Technical training details can be found in \cref{ch:appendix_implementation}.

% MLP is not the final solution, but can be extended
We find the MLP to be very effective on rest frame spectra on a fixed grid with full coverage, additionally to being computationally very efficient. Other architectures might be considered when addressing observational constraints such as a shifting viewing window. Here we make the assumption that we always have full access to the full spectral range in rest frame. We stress again that extensions seamlessly integrate into our framework, as the MLP AE can be swapped out for alternative architectures.

\subsubsection{Normalising Flow}
\label{ch:nf}

% normalising flow introduction
Normalising flows are a class of invertible neural networks for density estimation that map between two distributions. Usually they are trained to map from a complex data distribution to a simple distribution that can easily be evaluated and sampled from (e.g. a multivariate unit Gaussian). The flow is trained by minimizing the negative log-likelihood of the training data, which is given by the Gaussian likelihood of the transformed sample and the flow's Jacobian. Samples can then be drawn from a unit Gaussian and transformed back to data space via the inverse transformation. 

% our flow choice
We use a Masked Autoregressive Flow (MAF) \citep{papamakariosMaskedAutoregressiveFlow2018}, which is a standard choice that is well established for low-dimensional problems. 
Technical details on the flow implementation and training are listed in \cref{ch:appendix_implementation}.
As demonstrated in \ref{ch:flow_results}, the MAF is complex enough to achieve good results on the latent distribution. Our latent space is relatively simple, as most of the data complexity is absorbed in the AE. Again, we intentionally chose a simple model that can easily be substituted by another type of density estimation algorithm.

% adding z and norm to latent space
Because the AE was trained on normalised rest frame spectra, both redshift and luminosity are required to sample full observed-frame spectra. We add them to the model by appending them to the latent space, resulting in \num{8} dimensions in total. The flow models the distribution $p(\xi_i, z, L)$ with the 6 components $\xi_i$ of the AE latents, which represent normalised rest frame SEDs. Spectra are then sampled by sampling from this joint distribution, mapping the latents $\xi_i$ to spectra with the AE decoder, rescaling with the normalisation $L$, and redshifting with the sampled redshift $z$.

\subsection{Pseudo-Photometry}
\label{ch:photometry}

% pseudo-photometry
In order to evaluate model accuracy for photo-$z$ estimation, we calculate galaxy colours from their spectra. 
First, we transform the raw generated spectra from the AE decoder into observed-frame using the corresponding sampled redshifts and amplitudes from the flow. Then we simulate photometry from the spectra by integrating over passbands, both for the NN-generated and true simulated samples. We use KiDS-VIKING filters for compatibility with the DC3R2 SOM we use in \cref{ch:photo-z} \citep{wrightKiDS+VIKING450NewCombined2019, j.mcculloughDESICompleteCalibration2024}.

% photometric noise model from COSMOS levels
Additionally, we perturb the simulated photometry with measurement noise to reproduce the photometric scatter represented by the SOM. The DC3R2 SOM \citep{j.mcculloughDESICompleteCalibration2024} is based on the C3R2 SOM, which was trained on COSMOS data \citep{mastersCompleteCalibrationColor2017} and then remapped to KiDS-VIKING filters as detailed in \citet{j.mcculloughDESICompleteCalibration2024}. We therefore infer the noise level from the reported $3\sigma$ depths of the COSMOS15 catalogue \citep{laigleCOSMOS2015CATALOGEXPLORING2016} as these correspond to the colour scatter imprinted on the SOM through it's training data. 
We approximate the flux errors as background-dominated, independent Gaussian noise in each band.
We estimate uncertainties from the noisy fluxes and propagate the flux uncertainties to magnitudes and colours.
This is a simplified approximation (neglecting e.g. the mismatch in observing conditions, source-dependent noise, and error correlations) but is sufficient to obtain physically meaningful SOM assignments.
Finally, to emulate non-detections of certain photometric bands in the data, we label fluxes as upper limits when there is not at least a $3\sigma$ flux detection, i.e. $f_{\mathrm{band}} \lesssim 3\sigma_f$. This affects only u-band ($6\%$ of datapoints) and g-band ($\sim0.04\%$). We consider $3\sigma$ the upper limit on flux in this case.

\subsection{Photometric Redshifts}
\label{ch:photo-z}

% SOM for photo-z
For evaluating the impact on photo-$z$ estimation for cosmology, we perform a realistic photo-$z$ calibration. 
We choose a self-organising map (SOM)-based approach \citep{kohonenSelfOrganizingMaps2001}, which was adopted by both DES Y3 and Y6 \citep{mylesDarkEnergySurvey2021, yinDarkEnergySurvey2025}, as well as KiDS-1000 and KiDS Legacy \citep{hildebrandtKiDS1000CatalogueRedshift2021, wrightKiDSLegacyRedshiftDistributions2025} for their WL analyses.

% how does a SOM work
The SOM is an algorithm to map the variety of galaxy colours onto a 2-dimensional grid of cells.
It can be understood as binning of galaxies by similar colours, which was trained to preserve the structure of the colour-space spanned by the training data.
Galaxies are assigned to SOM cells according to which cell's colour weight vector they are most similar to under some distance metric. The best matching unit ($\mathrm{bmu}$) of a galaxy with colours $c$ and uncertainties $\sigma_c$ is calculated as 

\begin{equation}
    c_{\mathrm{bmu}} = \argmin_{c_i \in c_\mathrm{SOM}} \Bigg|\frac{(c - c_i)^2}{\sigma_c ^2}\Bigg| 
\end{equation}

As a result, all galaxies in a bin share similar colours and are assumed to have similar properties. In a given cell, this produces a redshift posterior $p(z|\vec{c})$. 

% we choose a fixed, pre-trained SOM
For this work we adopt the pre-trained DC3R2 SOM \citep{mastersCompleteCalibrationColor2017, j.mcculloughDESICompleteCalibration2024}.
While the colour space mapped by this SOM is not tailored to our data, and thus will not yield optimal photo-$z$ performance, we use it as a fixed and realistic prescription for tomographically binning the data.

% inherent population mismatch in SOM 
We note that there is an intrinsic mismatch between the SOM colour space and our training data due to representing inherently different galaxy populations. Data selection, redshift- and magnitude-limits, photometric noise properties, and the underlying galaxy population differ between COSMOS data and our \textsc{GalSBI-SPS} simulations. 
This is partly because \textsc{GalSBI-SPS} has not yet been constrained against data of COSMOS-depth. Parameters were chosen instead from literature studies from GAMA and DEVILS, hence it is more tailored to represent a shallower galaxy population.
This is reflected in lacking cell coverage when sorting galaxies into the SOM (see \cref{ch:photo-z_results}), but does not affect the results of our study as long as the overlap is significant enough to achieve a reasonably resolved, physically meaningful binning.
In this work, we do not attempt to reconcile \textsc{GalSBI-SPS} with COSMOS data. Instead we merely aim to present a schematically accurate pipeline to compare the impact of generated vs. simulated samples.

% distance metric is \chi^2, limit handling
Following the prescription by \citep{mastersCompleteCalibrationColor2017}, the distance metric is a reduced $\chi^2$ based on the calculated colours and colour-uncertainties. In the case of non-detections, i.e. when only an upper flux limit exists in a band, the colour only enters the distance when the bound is violated.

% tomographic bins
We sort the simulated sample into the SOM, and evaluate the mean redshift in each cell. We sort cells by redshift and based on these split the data into 5 equally populated tomographic bins, following \citet{j.mcculloughDESICompleteCalibration2024}. 
SOM cells are assigned to bins by progressively choosing cells with the lowest $\langle z \rangle$, until the number of galaxies in each bin is greater than one fifth the total number of galaxies. We keep this cell-to-bin assignment fixed, and use it also for constructing tomographic bins for the PAE sample. 
Because bin-assignment is not well defined when a cell is empty, we omit cells which are not populated by both samples. This is mostly affecting cells with very low galaxy count where shot-noise, photometric scatter, or a slight mismatch in the tails of the distribution dominate (see \cref{ch:photo-z_results}).

% reweighting
In a typical galaxy survey application, a photometric sample is calibrated by a smaller sub-sample with confident redshifts. Comparing the SOM cell coverage and occupancy in each cell allows the diagnosis and correction of selection effects. The spectroscopic sample can be reweighted by the relative abundances in a cell to match the photometric sample. In practice this allows more efficiency with the calibration data, as a representative sample with confident redshifts is only required on a per-cell level.
We employ this reweighting as well, to match the abundances better and to correct for some of the mismatch between the populations (see \cref{ch:photo-z_results}). We calculate the redshift distribution of a selection of cells $\mathcal{C}$ as 
\begin{equation}
    \label{eq:nofz_weighting}
    n^\mathcal{C}_\mathrm{PAE}(z) = \sum_{i \in \mathcal{C}} \frac{N^i_{\mathrm{sim}}}{N^i_{\mathrm{PAE}}} n^i_{\mathrm{PAE}}(z) , 
\end{equation}

where $N^i$ are the abundances in cell $i$ and $n^i(z)$ is the redshift distribution in cell $i$. 
The subscript $\mathrm{PAE}$ indicates the dataset generated by the PAE, and $\mathrm{sim}$ refers to the test-set simulated with \textsc{GalSBI-SPS}, i.e. a disjoint sample of $10^6$ spectra drawn from the same distribution as the PAE training data, representing the target distribution we want to match.
This mirrors the approach applied in real data, where the simulations take the role of the photometric sample and the PAE corresponds to the spectroscopic calibration sample.
In \cref{ch:photo-z_results} we also quote the results without the reweighting applied, which slightly weakens the agreement.

%% file: sections/3Results.tex
Here, we inspect the results of the trained model. \Cref{ch:ae_results} and \cref{ch:flow_results} show the performance of the autoencoder and normalising flow respectively. \Cref{ch:colour_results} inspects the learned distribution of galaxy SEDs on the population level. We evaluate the impact of our model on photometric redshifts in \cref{ch:photo-z_results}.

\subsection{Autoencoder reconstructions}
\label{ch:ae_results}

% AE loss distributions
\begin{figure}[ht]
    \centering
    \includegraphics{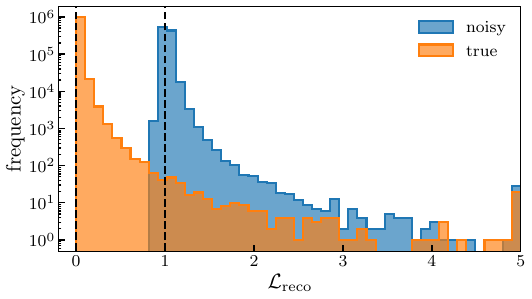}
    \caption{Reconstruction loss between AE denoised spectra, and noisy and noiseless spectra respectively. The histograms show the distributions of the test-set, with the last bin containing overflows. The loss when comparing to data is effectively bounded from below by 1 due to the inherent scatter of the inputs. The noiseless losses are significantly smaller, with $97\%$ of points achieving a loss $<0.1$.}
    \label{fig:AEloss}
\end{figure}

% reconstruction loss interpretation
We train the autoencoder on simulated spectra from \textsc{GalSBI-SPS} with added Gaussian noise as described in \cref{ch:ae} and evaluate performance it's performance on the unseen test-set. We can inspect the AE reconstructions in two key ways: 1. comparing the reconstructed spectra to the inputs, i.e. how the NN is trained, and 2. comparing how close the reconstructions are to the underlying ground truth, which is available only in simulations.
Since the reconstruction loss is the average MSE divided by the true data variance, there is a physical interpretation to its numerical value: Optimal reconstruction with perfect denoising would achieve $\mathcal{L}_{\mathrm{reco}}(\text{\AE}(\tilde{f_\lambda}), \tilde{f_\lambda}) = \mathcal{L}_{\mathrm{reco}}(f_\lambda, \tilde{f_\lambda}) = 1$ on noisy data, given by the inherent scatter of the data around the ground-truth. 
This is also a lower bound on the achievable precision: values $<1$ would indicate over-fitting.

% reconstruction loss values and metrics
Our model achieves $\mathcal{L}_{reco} = 1.02$ on the noisy test set, indicating near-optimal performance. 
When comparing de-noised spectra to the noiseless simulations, the AE achieves $\mathcal{L}_\mathrm{reco} = 0.03$, demonstrating that the denoising successfully reconstructs spectra that are remarkably close to the ground-truth.
The per-spectrum $\mathcal{L}_{reco}$ value distributions, comparing to both noisy and ground-truth spectra, are shown in \cref{fig:AEloss}. 
The mean absolute deviation is $\Bigl \langle \frac{|\text{\AE}(\tilde{f_{\lambda,i}}) - f_{\lambda,i}|}{\sigma_{\lambda,i}} \Bigr \rangle_{\lambda,i} = 0.1$, which means that the reconstructions are on average $0.1 \sigma$ away from the truth.
This corresponds to a mean relative flux error of $1\%$, averaged over the dataset and pixels.
By integrating the reconstructed spectra over photometric filters, we find that reconstructed magnitudes scatter with a standard deviation of $0.01$ mags around their true values.

% per-pixel residual distributions
\begin{figure}[ht]
    \centering
    \includegraphics{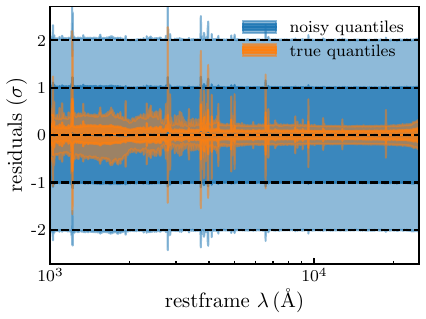}
    \caption{
    Distribution of autoencoder reconstruction residuals $f_\lambda-\text{\AE}(\tilde{f_{\lambda}})$ in units of $\sigma_\lambda$, resolved by wavelength bin, as evaluated on the test-set. The shaded bands represent the $68\%$ and $95\%$ quantiles, while the central line shows the median. The noisy data scatter is close to Gaussian around the reconstructions, which is reflected by the quantiles matching the expected width. The residuals with respect to the ground truth are much smaller, well below the $1\sigma$ region almost everywhere, and not visibly biased. The deviations are significantly larger at emission lines.}
    \label{fig:AEresiduals}
\end{figure}

% per-pixel residuals
The per-pixel residuals shown in \cref{fig:AEresiduals} give some insight into at which parts of the spectrum the AE is most accurate. 
The deviations with respect to the inputs follows the expectation from Gaussian scatter, while the deviations to the truth are very closely confined around zero, which is consistent with the averaged reconstruction error shown in \cref{fig:AEloss}. 
Overall, the residual distribution is largely unbiased. The AE seems to struggle most with reconstructing emission lines precisely, which is expected since our loss function weights every pixel equally and thus does not put special emphasis on the lines, but tends to focus on the continuum.

% reconstruction example
\begin{figure*}[ht]
    \centering
    \begin{subfigure}[b]{\textwidth}
        \centering
        \includegraphics[]{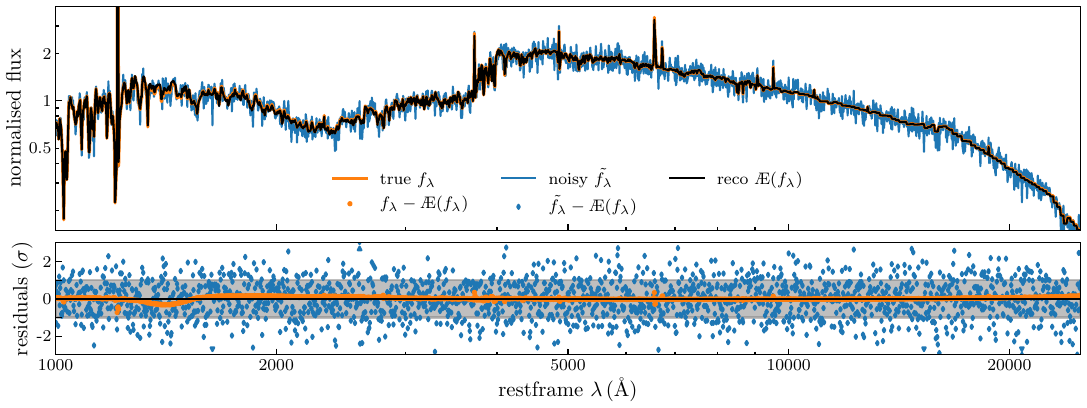}
        % \caption{}
        % \label{fig:AEreco_ex}
    \end{subfigure}
    \begin{subfigure}[b]{\textwidth}
        \includegraphics[]{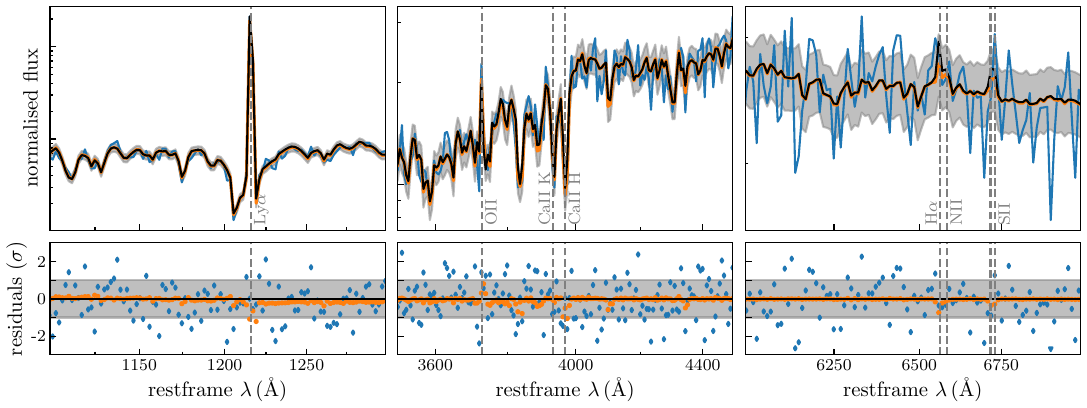}
        % \caption{}
        % \label{fig:AEreco_zooms}
     \end{subfigure}
     \caption{Exemplary autoencoder denoising on the test-set. The noisy input spectrum is compressed by the AE and reconstructed. The denoised spectrum is very close to the ground-truth, with residuals mostly $\ll 1$. The upper panel shows the full spectrum reconstruction of an exemplary galaxy. The lower panel shows zoom-ins of the reconstruction of a second spectrum from the test-set. Regions shown focus on the $\text{Ly}\alpha$ line, \qty{4000}{\angstrom}-break, and $\text{H}\alpha$ line regions. The $\text{Ly}\alpha$ line has residuals significantly larger than uncertainties, which shows that the AE struggles most with reconstructing emission lines. The recovery of the $\text{H}\alpha + \text{NII}$ and $\text{SII}$ lines demonstrates that the AE is able to resolve and recover small-scale shapes of the spectrum, even when they are below the noise level individually.}
     \label{fig:AEreco}
\end{figure*}

% exemplary reconstruction
\Cref{fig:AEreco} shows an example of a typical AE reconstruction. The reconstructed spectrum follows the true spectrum closely, proving that the denoising is effective and meaningful in that it gets close to the true spectra. While imprecise predictions are observed at significant emission lines, it is accurately capturing the small scale shape of the true spectrum, even below the noise level.

% AE denoising performance summary
Overall the AE effectively denoises the inputs, and thus models noiseless spectra, while learning only from noisy data. The network has never seen the ground-truth during training, and is still able to recover realistic spectra that are close to the simulations. At the same time it drastically compresses spectra down to 6 dimensions without relying on any spectral templates. We note that the performance shown here is under the assumption of a simplified Gaussian noise model. We discuss the implications of the more complex noise in real data in \cref{ch:discussion}.

\subsection{Flow Latent Space}
\label{ch:flow_results}

% flow learned distribution
\begin{figure*}[ht]
    \centering
    \includegraphics[]{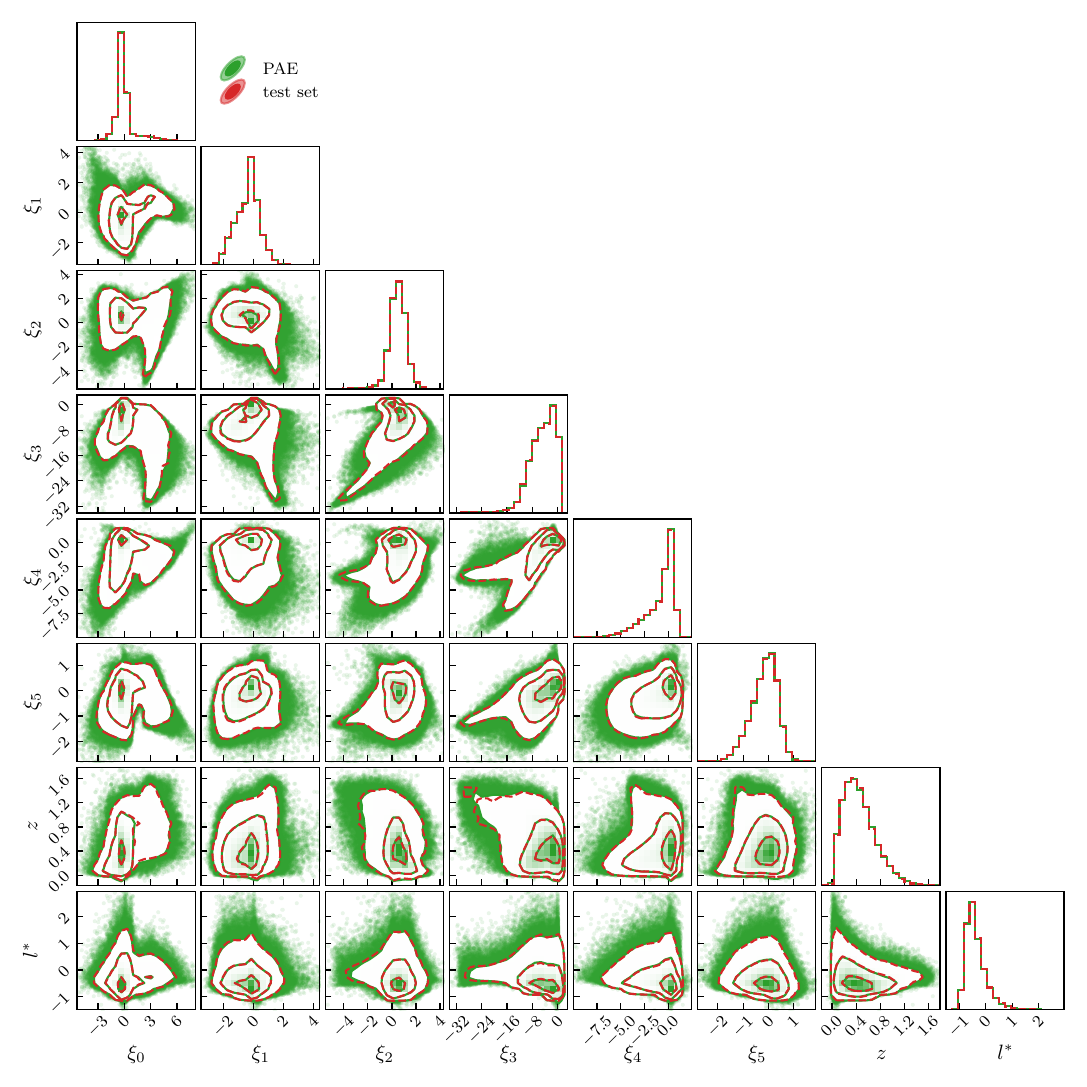}
    \caption{The distributions of the latent space of simulated and generated samples. The solid and dashed contours show the $1$, $2$, and $3 \sigma$ levels. The flow's contours are shown in green, and reproduce the target distribution in red well, with deviations visible only at the $3\sigma$-level.}
    \label{fig:latentdist}
\end{figure*}

%flow results
We append the redshift $z$ and normalisation $L^*$ to the latent space of the AE, where we rescale $L^*$ with the fixed transformation $l^* = \log(L^*\cdot10^{17})$, and train a normalising flow to learn the joint pdf $p(\xi_i,z,l^*)$, as described in \cref{ch:nf}.
\Cref{fig:latentdist} shows the distribution of both the latent space of the simulations, and samples from the flow. They overlap very well, with deviations only visible at the $3\sigma$ contours.
The latent space is not explicitly regularised during AE training, still it shows a relatively simple (localised and largely unimodal) distribution, which is advantageous for modelling with a simple MAF.

\subsection{Matching sampled distribution}
\label{ch:colour_results}

% z-scores
\begin{figure}[ht]
    \centering
    \includegraphics{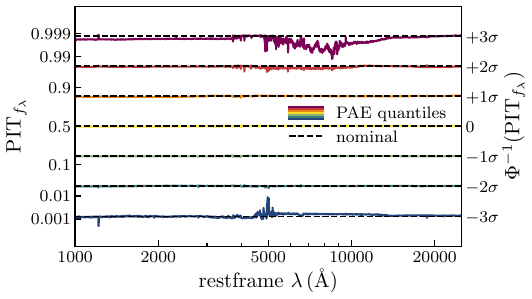}
    \caption{Pixel-wise PIT under the target distribution, applied to PAE quantiles. Coloured lines show where the $1$, $2$, and $3\sigma$ PAE quantiles and median fall relative to the simulation quantiles indicated by dashed horizontal lines. The y-axis is rescaled with a probit-transform for better visibility of the outer quantiles.
    The samples are generally consistent and show that out model is unbiased. Noisy deviations are visible only for the outermost quantiles, which implies a slight disagreement in the very tails of the distributions.}
    \label{fig:AE_flow_zscore}
\end{figure}

% pixel-wise z-scores
We evaluate the joint performance of both networks by investigating the distributions of each $10^6$ spectra generated by the PAE and from the test-set.
We compare the overall pixel-wise marginal distributions of both samples in flux space in \cref{fig:AE_flow_zscore}, which shows the probability integral transform (PIT) of PAE sample quantiles under the target distribution. 
The quantiles of identically distributed samples would by definition be represented by horizontal lines in this plot. Deviations indicate that the PAE predicted quantile is above or below the target quantile.
In this case, the PAE predictions are remarkably consistent with the simulations, with visible differences mostly only in the $3\sigma$ quantiles, which probes the low-statistics tails of the distributions.
As this probes only the per-pixel marginal distributions, and not the shapes of individual SEDs, we also show the derived colour distributions in \cref{fig:colordists} which show similar agreement as previous results.

\subsection{Impact on photo-z calibration}
\label{ch:photo-z_results}

% SOM filling
We sort galaxies by colours into the SOM, according to the prescription outlined in \cref{ch:photo-z}.
With the \textsc{GalSBI-SPS} sample we fill $31\%$ of the cells, which goes up to $40\%$ when only considering cells whose C3R2 nominal redshift is compatible with our redshift cut $z_{\mathrm{max}} = 1.55$. Of these, more than half are populated with multiplicity $>10$. As pointed out in \cref{ch:photo-z}, this can be attributed to the inherent population mismatch between the SOM and \textsc{GalSBI-SPS}, but still serves as a physically meaningful binning by colour. 
See also \cref{fig:SOM_occupancy} for the regions of the SOM colour-space filled by the simulations. Overall, $\sim3000$ SOM cells are populated by both the simulated and learned samples, indicating good overlap in colour space, with $\sim 400$ cells each being filled by one sample but not the other. The latter affects mostly cells with a very low number of galaxies, and can thus be attributed partly to Poisson noise in the tails of the distributions. As implied by \cref{eq:nofz_weighting}, they are implicitly discarded for calculating redshift distributions $n(z)$.

% mean z differences
\begin{figure}
    \centering
    \includegraphics[]{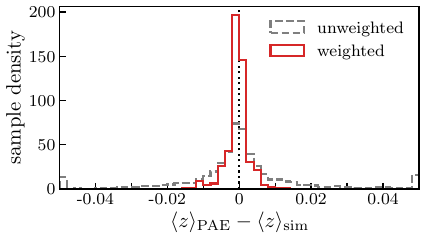}
    \caption{Histogram of mean redshift deviations per cell between the simulated and generated samples. The first and last bins contain outliers. The black histogram shows the per-cell differences in mean redshift derived from the simulation and the PAE. The red histogram weights each cell by galaxy-count, which effectively corresponds to the per-galaxy difference in photo-$z$ labels.
    Deviations are very tightly confined around $0$, with few extreme outliers which are down-weighted due to small number statistics.}
    \label{fig:photozpercelldeviations}
\end{figure}

% cell-wise mean z differences.
\Cref{fig:photozpercelldeviations} shows the differences of mean $z$ between both approaches, both per-cell and weighted by the abundance of test galaxies in a cell. The latter corresponds to the distribution of the difference in photo-$z$ prior per galaxy. Deviations are typically very small, and scatter tightly around zero, with negligible bias. The significant deviations observed in the per-cell distribution can be attributed to cells with low number-statistics. 

% n(z)
\begin{figure*}
    \centering
    \includegraphics[]{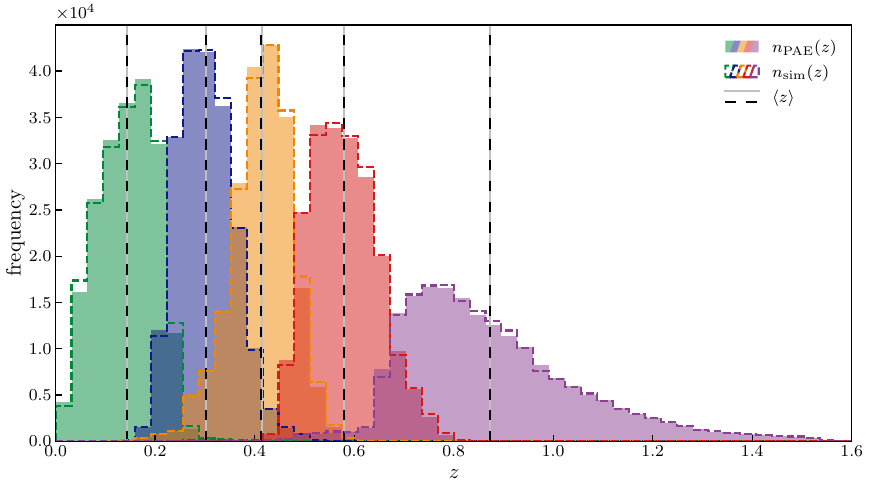}
    \caption{Tomographic $n(z)$ distributions for 5 tomographic bins, both for calibration with the simulated and generated samples. The mean redshifts for every bin are indicated by vertical lines: solid gray lines representing the PAE sample, and dashed black lines for the test sample. They overlap very well with deviations $ | \langle z^{\text{phot}}_{\text{PAE}} \rangle - \langle z^{\text{phot}}_{\text{true}} \rangle | \leq 0.0007 (1 + z^{\text{phot}}_{\text{true}})$}
    \label{fig:nofztomographic}
\end{figure*}

% per-bin deviation table
\begin{table}[ht]
    \centering
    \caption{Mean redshifts of the populations inside the five tomographic bins. $\langle z_\mathrm{sim} \rangle$ is derived from $10^6$ test-set spectra, while $\langle z_\mathrm{PAE} \rangle$ is calculated with $10^8$ PAE samples to suppress stochastic scatter.}
    \sisetup{input-comparators = {<=>\approx\ge\geq\gg\le\leq\ll\sim\lesssim}}
    \begin{tabular}{c
        S[table-format=1.4]
        S[table-format=1.4]
        S[table-format=>-1.5]}
        \toprule
        {bin} & {$\langle z_{\mathrm{sim}} \rangle$} & {$\langle z_{\mathrm{PAE}} \rangle$} & {$\Delta \langle z \rangle \, / \, ( 1 + \langle z \rangle )$} \\
        \midrule
        1 & 0.1438 & 0.1434 & -0.00035 \\
        2 & 0.3018 & 0.3017 & -0.00014 \\
        3 & 0.4142 & 0.4132 & -0.00069 \\
        4 & 0.5796 & 0.5795 & -0.00005 \\
        5 & 0.8736 & 0.8745 &  0.00045 \\
        \addlinespace
        \multicolumn{3}{r}{All bins absolute bias} & < 0.0007 \\
        \multicolumn{3}{r}{Stage-IV requirements} & < 0.001 \\
        \bottomrule
    \end{tabular}
    \label{tab:summary}
\end{table}

% n(z) statistics
Finally, we construct colour-based tomographic bins, as described in \cref{ch:photo-z}.
We find that with $10^6$ galaxies, the PAE model is so accurate that sample variance per tomographic bin becomes non-negligible. Because sampling from the PAE is very fast, we can repeatedly sample new spectra and investigate the variance on mean redshifts. From $100$ repeated runs, we estimate a scatter of up to $\sigma_{\langle z_\mathrm{PAE} \rangle} \lesssim 0.00009\, (1 + \mu_{\langle z_\mathrm{PAE} \rangle})$. By aggregating these runs, we can suppress the PAE sample scatter to get a closer estimate of the systematic bias induced by the trained model. The resulting $\Delta \langle z \rangle$ does still include two sources of sample variance: the limited size of the test-set, for which it is more expensive to re-run the simulations repeatedly, and the sample variance imprinted on the model through it's training data, which is irreducibly limited by how many spectra we can realistically obtain with multiplexed surveys.

% n(z)
The tomographic redshift distributions for both the simulation and PAE samples are displayed in \cref{fig:nofztomographic}. The $n(z)$ agree very closely. While slight deviations of the histograms are visible, the mean redshift of each bin agrees very well. The exact values of $\langle z \rangle_\mathrm{bin}$ can be found in \cref{tab:summary}. For each bin, the difference is $ | \langle z \rangle ^{\text{PAE}} - \langle z \rangle ^{\text{true}} | < 0.0007 (1 + \langle z \rangle ^{\text{true}})$, which meets the requirements for Stage IV surveys of $| \delta \langle z \rangle | \leq 0.001(1+z)$.
These results are slightly weaker when considering the raw population, that is without the per-cell reweighting procedure described in \cref{ch:photo-z}, at $ | \langle z \rangle ^{\text{PAE}} - \langle z \rangle ^{\text{true}} | < 0.0013 (1 + \langle z \rangle ^{\text{true}})$.

%% file: sections/4Discussion.tex
These results demonstrate that generative models can potentially learn a galaxy population model that meets the Stage-IV photometric redshift calibration accuracy requirements. We evaluated the systematic bias that the learned prior induces on a cosmological analysis under a SOM based photo-$z$ calibration, which is consistently below Stage-IV required accuracy for every bin.
We demonstrated this on a set of mock spectra from the \textsc{GalSBI-SPS} population model, which allowed us to compare the results to the underlying ground-truth. 

The prior represented by the NN can be used in any inference scheme and can interpolate the training data, and effectively bootstrap existing spectroscopic samples. 
The generative model itself does not rely on exterior information and just learns from a set of SEDs, and can thus be trained without making any assumption on the galaxy population when spectroscopic training data is available.
While this study was performed under several idealising assumptions, which we discuss below, the implication is that generative models are a promising way forward towards precision photo-$z$ characterisation. 

% assuming representative sample
Throughout this study, we neglected to factor in selection effects, which are a major hurdle in practice. The spectroscopic sample is in general not representative of the photometric sample, which will bias the results. 
We implicitly assumed that the calibration sample is representative, addressing biases only insofar as we re-weight the $n(z)$ distributions by the relative galaxy counts in each colour-cell. Moreover, in this work we did not address that photometric weak-lensing samples are in general substantially fainter than the spectroscopic calibration data, which we leave to future work.

% but we are doing not worse than anyone else
However, without assuming a physical model the fainter population remains fundamentally inaccessible and this issue can only be addressed by large spectroscopic surveys and investment of long exposure times, such as it is the case for DESI and 4MOST. 
Moreover, \citet{schmidtEvaluationProbabilisticPhotometric2020} evaluated discriminative photo-$z$ under the same assumption of representative samples, and found that no method was able to achieve Stage-IV level. 
We improve on these results by building a model for rest-frame spectra: because the redshift transformation is made explicitly, a single SED type observed at any redshift informs predictions at every other redshift, which improves robustness at poorly populated regions at higher redshift.
A natural test of this robustness, which we leave to future work, is to repeat the analysis under a Rubin-like selection rather than the 4C3R2-like spectroscopic sample used here, and to progressively impose more magnitude cuts on the spectroscopic training set to quantify how performance degrades as representativeness decreases.

% noise model
One major simplification compared to real data is our noise model. We assumed Gaussian noise with a constant amplitude across the dataset. 
Real observations are more complicated, with strongly heteroscedastic noise across both the dataset, as well as across wavelengths for each individual spectrum. 
The strongly heterogeneous noise is likely to be a non-trivial challenge when training on real data.
Additionally, noise in real spectra can be correlated due to detector effects, observational systematics, and e.g. sky lines.
The excellent performance we observed for the AE is partly explainable by the Gaussian assumption: the loss function is effectively equivalent to the true likelihood of the problem.
Full forward modelling of all systematic effects can potentially address this issue, but it is not obvious how the net performance holds up in a realistic scenario.

% wavelength coverage
Furthermore, we assumed full access to the entire rest-frame spectrum from the UV to infrared. We made this simplification because the pre-trained SOM we used operates on KiDS-Viking photometry, i.e. ugrizYJHKs-bands.
In observations however, only a sliding window of each spectrum is accessible due to the limited wavelength coverage of our detectors and galaxies being redshifted. Our naive MLP is not ideally suited to deal with this type of data, but other architectures, such as CNNs, transformers, or masking inputs like SPENDER \citep{melchiorAutoencodingGalaxySpectra2023}, could help address this practical limitation. 
Moreover, in real data, redshift failure outliers complicate rest-frame SED modelling when the rest-frame is erroneously shifted for a small fraction of the training dataset.

% pseudo photometry
Finally, we simulated photometry by integrating the SEDs, which implicitly assumes that colours and spectra are measured in the same aperture, and that the photometric calibration is perfect, which does not hold exactly in practice. In a realistic survey, the fibre aperture probes a different region than the photometry measured in an image, and calibrations are imperfect, both of which adds additional scatter to the relation between colours and SEDs.

% possible extensions
As outlined before, our approach is modular, which implies that several of the above assumptions can possibly be alleviated by upgrading parts of the architecture. 
Moreover, the density estimation, which is a crucial part that requires exact calibration, can be upgraded by implementing recent state-of-the-art density estimation techniques such as neural spline flows, continuous normalising flows, or diffusion models, if the complexity of the latent space demands it. 
We chose a simple rendition of this type of PAE, demonstrating that generative models are expressive enough to learn priors to match the colour-redshift relation precisely given enough training data.

%% file: sections/5Conclusion.tex
% extablising probem
In this paper we investigated the application of generative modelling in the context of photometric redshifts for Stage-IV galaxy surveys. 
Photo-$z$ inference -- implicitly or explicitly -- relies on a prior of the galaxy population, that is a pdf of galaxy SEDs and how they evolve with redshift. Existing models fail to reach the projected LSST and Euclid WL requirements of $\langle z \rangle < 0.001(1+z)$, even under idealising assumptions.

% genAI can help
We concluded that in order to improve photo-$z$ inference, a prior model would need to not rely on templates or  assumptions of any kind and incorporate a learned model of intrinsic rest-frame galaxy SEDs.
We apply a machine learning generative model as an instance of this prior, learning the distribution of galaxy spectra. 

% validating on simulations
To enable robust comparisons to the underlying ground-truth, we simulate a realistic set of mock spectra with the \textsc{GalSBI-SPS} galaxy population code. We train the model on this simulated dataset with added Gaussian noise to show that we accurately recover the underlying population. Because the NN is agnostic to what the data looks like, we can expect the results to generalise to training on observations if the simulations are realistic enough, even when they represent a different population.

% trained PAE
We trained a probabilistic autoencoder that consists of a first dimensionality reduction step with an AE followed by a latent density estimation with a normalising flow.
We demonstrate that the AE can effectively denoise the inputs and find 6 dimensional representations of noiseless spectra, facilitating precise density estimation. We find excellent agreement between NN outputs and the simulations, both on the level of individual SED reconstructions, as well as on the population level distributions.

% impact on photo-z
Finally, our metric of success is the impact of the training in the context of a realistic photo-$z$ calibration. We employ the C3R2 SOM to assign galaxies to tomographic bins based on colours and calculate redshift distributions. 
We find that the bias on the mean redshift is $\Delta \langle z \rangle \leq 0.0007 (1+z)$ in each tomographic bin, which is compatible with the per-mille requirements of Stage-IV. Though this is for a brighter subset of galaxies than Stage-IV surveys will observe and assuming representative spectroscopic coverage for the scope of this paper.

% modular and flexible approach with improvements
Our approach is flexible, modular, and easily adaptable. Improvements to the model architecture can help address observational effects and limitations. 
Further validation on more realistic yet controlled conditions is required in order to build trust in the model to rely on it when trained purely on data with no true labels.

% proof-of-concept for genAI
We see this as a proof-of-concept that generative models are a feasible avenue for modelling the galaxy population and enabling photo-$z$ measurements at the Stage-IV level to attain precise, robust cosmological inference.

%% file: sections/Appendix.tex
\section{Implementation}
\label{ch:appendix_implementation}

In the following we describe technical details of the NN training.
The code produced for this paper is publicly available on GitHub\footnote{\url{https://github.com/nfrediani/PAE-galaxy-SEDs}}.

\subsection{Autoencoder architecture \& training}
\label{ch:appendix_implementation_ae}

% AE architecture
The autoencoder architecture consists of fully-connected layers with hidden layer sizes of [1000, 500, 100, 6] and ReLU activation. The Decoder is a mirrored copy of the Encoder architecture.

% asinh for normalisation
Additionally for preprocessing the inputs, we prepend the network with a fixed $\mathrm{asinh}(x)$ transformation. Correspondingly outputs are transformed back to flux space by $\mathrm{sinh}(\hat{x})$ before evaluating the loss function. This allows us to still optimise in flux-space, which avoids Jensen's inequality bias, while still benefiting from the numerical advantages of range compression for the NN. This is especially relevant because the inputs span a wide dynamic range. To increase numerical stability, we employ gradient clipping at a grad norm of $1.0$.

% AE training
We train using the Adam optimiser with a learning rate of $10^{-3}$ for 100 epochs. If the loss does not improve for 5 epochs, the learning rate is reduced by a factor of 2. We checkpoint the model after each epoch and use the stage with the lowest validation loss for downstream tasks.

% optimisation disclaimer
We did not perform systematic hyperparameter optimisation, but arrived at this configuration through manual experimentation.
Extensive optimisation would potentially lead to better performance, but given our results, gains would be marginal and insubstantial for the purpose of this study.

\subsection{Flow architecture \& training}
\label{ch:appendix_implementaion_flow}

% flow architecture
We implement the MAF using \texttt{zuko}\footnote{\url{https://github.com/probabilists/zuko}}, with $8$ transformations and random orderings for each layer.

% flow training
We train the flow with the Adam optimizer with a constant learning rate of $10^{-3}$ for 100 epochs. If the loss does not improve for 5 epochs, the learning rate is reduced by a factor of 2. We save the best-performing model on the validation set.

% optimisation disclaimer
Similarly to the AE, we did not optimize the flow architecture and training setup by performing formal hyperparameter search, but instead manually experimented with different setups.
The final performance of the density estimation could potentially be improved by more rigorously exploring model choices. This includes more advanced density estimation techniques such as flow matching or diffusion. For this study however, we deemed further performance improvement to be marginal and stuck with the simplest option that still achieved good results.

\section{Autoencoder scaling with noise level and latent dimensionality}
\label{ch:appendix_scaling}

% scaling
In this section we show how the AE performance scales with the dimensionality of the latent space and noise level of the data by varying both parameters separately and training a NN from scratch. The rest of the hyperparameters are kept at their fiducial valued described in \cref{ch:ae} and \cref{ch:appendix_implementation_ae}.

\begin{figure}
    \centering
    \includegraphics[]{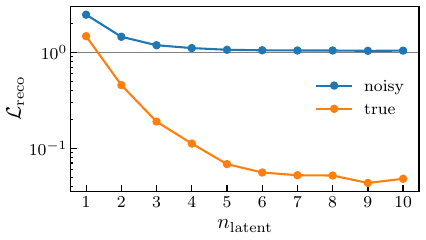}
    \caption{The scaling of autoencoder reconstructions with the size of the AE latent space. Shown are both the training loss in blue and the comparison to the ground-truth in orange. Both quantities improve with increasing latent size until they plateau around $n_\mathrm{latent} \approx 6$, which we adopt for our fiducial model.}
    \label{fig:AE_scaing_latent}
\end{figure}

% latent scaling
\Cref{fig:AE_scaing_latent} shows the test reconstruction loss for both noisy and noiseless spectra. As expected, reconstructions become more accurate with a larger latent space. We adopt $n_\mathrm{latent} = 6$ as our fiducial architecture, as increasing the latent dimension further gives diminishing returns.

\begin{figure}
    \centering
    \includegraphics[]{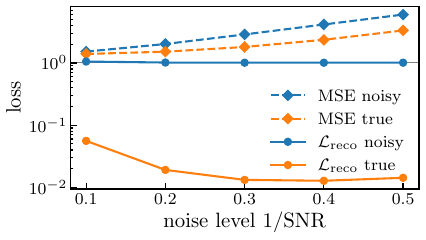}
    \caption{The scaling of autoencoder reconstructions with the noise level $\Sigma$. Shown are both the training loss in blue and the comparison to the ground-truth in orange, as well as a raw MSE in dashed for both values. The training loss stays constant at $\sim 1$, while the true loss actually improves in noisier scenarios. This is because $\mathcal{L}_\mathrm{reco}$ is weighting the absolute deviation by inverse variance. Reconstructions become relatively more accurate at higher SNR, while the absolute deviation increases as expected, as demonstrated by the raw MSE. For simplicity, we adopt a noise level of 0.1 throughout this work.}
    \label{fig:AE_scaling_noise}
\end{figure}

% noise scaling
\Cref{fig:AE_scaling_noise} shows how AE reconstructions scale with increasing the noise level $\Sigma$ of our Gaussian noise model $\tilde{f_\lambda} = f_\lambda \cdot (1 + \mathcal{G}(0,\Sigma))$. The noisy test loss is constant at $\approx 1$ by definition for a properly converged network as explained in \cref{ch:ae_results}. The comparison to ground-truth spectra actually becomes better for lower SNR. This is due to $\mathcal{L}_\mathrm{reco}$ being defined as MSE weighted by inverse variance. The absolute reconstruction errors do increase as more information is washed out by a higher noise level, as confirmed by the monotonically increasing MSE. However, AE performance degrades slower-than-linear with increasing noise level at high SNR, leading to better reconstructions relatively.

\section{PAE colour distributions}
\label{ch:appendix_colordists}

% colour distributions
\begin{figure*}
    \centering
    \includegraphics[]{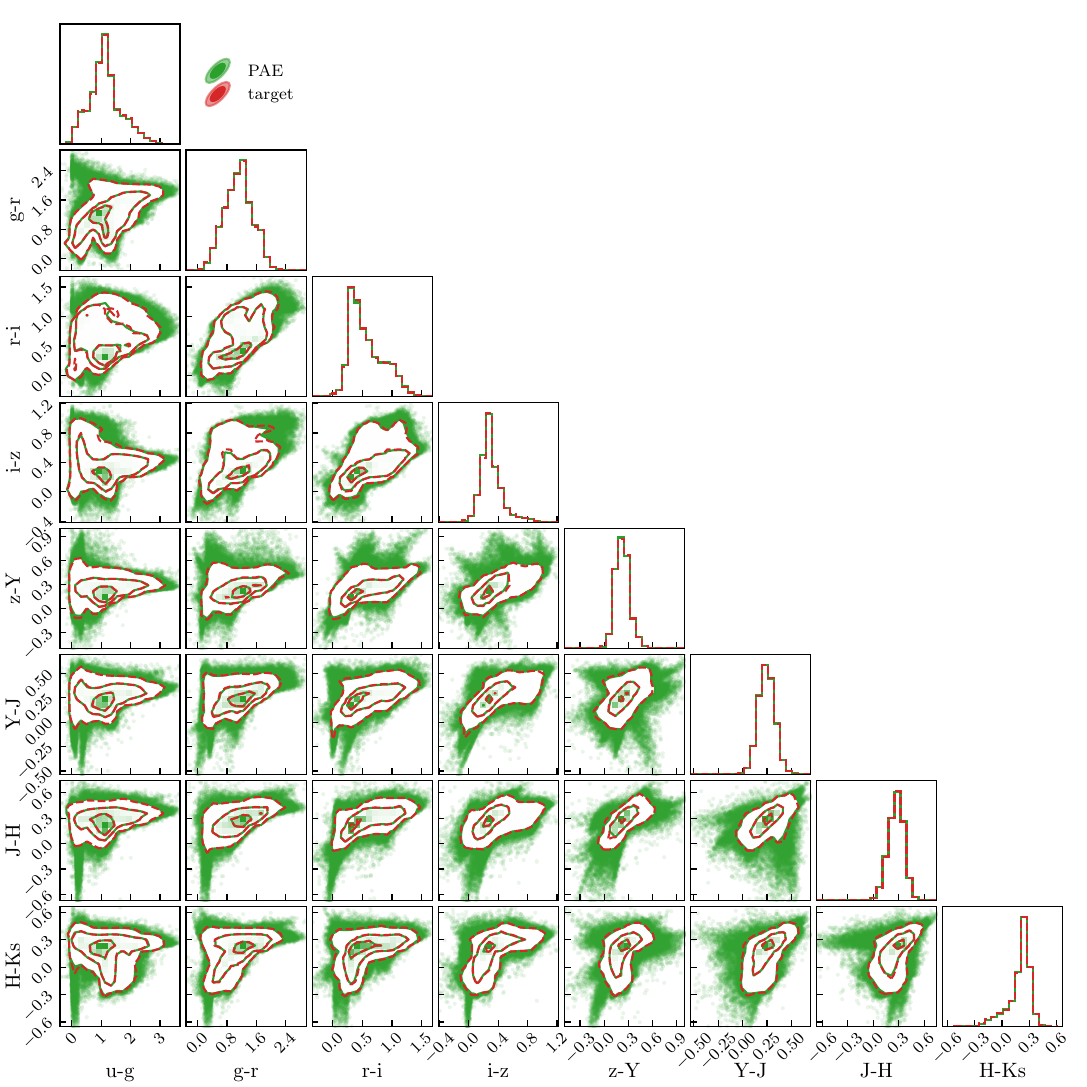}
    \caption{The colour distributions of the training sample in red and generated sample in green. The two distributions overlap very well, with some small deviations mostly at the $3\sigma$-level.}
    \label{fig:colordists}
\end{figure*}

% colour distributions
We simulate photometry and colours for both the simulated and PAE samples as outlined in \cref{ch:photometry}. Both colour-distributions are shown in \cref{fig:colordists}. They overlap very well up to $3\sigma$-levels. This is consistent with the performances of both individual networks of the PAE, but evaluates their accuracy jointly in physical space.

\section{SOM assignment}
\label{ch:appendix_som}

\begin{figure}
    \centering
    \includegraphics[]{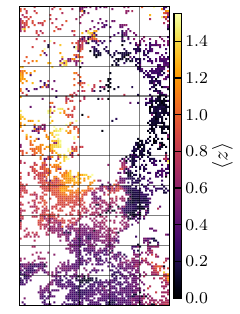}
    \includegraphics[]{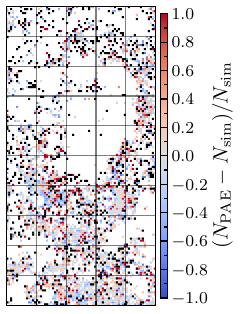}
    \caption{SOM occupation when assigning galaxy spectra to SOM cells. 
    The left panel shows the mean redshift of PAE spectra in every cell. The visible trends are in agreement with other versions of this SOM \citep{mastersCompleteCalibrationColor2017,j.mcculloughDESICompleteCalibration2024}.
    The right panel shows the relative deviation between the number of PAE samples and test samples in every cell. Black pixels indicate cells populated by either one of the samples but not the other. While there can be significant deviations on a per-cell basis, there are no obvious systematic trends.}
    \label{fig:SOM_occupancy}
\end{figure}

% SOM occupation
\Cref{fig:SOM_occupancy} shows the SOM assignment of galaxies and the mean redshift across colour space as represented by the SOM. With adding slight COSMOS-like photometric noise we populate $\sim31\%$ of cells. The filling fraction is limited by the inherent discrepancy between the colour spaces spanned by the simulations and the SOM weight vectors, which is primarily driven by the different selections, notably our brighter cut of $m_z < 22$. Additionally, some regions of colour space are populated by high redshift galaxies beyond our $z_\mathrm{max}=1.55$ cut \citep[c.f.][]{j.mcculloughDESICompleteCalibration2024}, so our sample is naturally not expected to populate those.